\documentclass[12pt]{article}

\usepackage{epsf,amsfonts,hyperref}
\renewcommand{\appendix}[1]{
    \addtocounter{section}{1}
    \setcounter{equation}{0}
    \renewcommand{\thesection}{\Alph{section}}
    \section*{Appendix \thesection\protect\indent #1}
    \addcontentsline{toc}{section}{Appendix \thesection\ \ \ #1}
}
\newcommand\encadremath[1]{\vbox{\hrule\hbox{\vrule\kern8pt
\vbox{\kern8pt \hbox{$\displaystyle #1$}\kern8pt}
\kern8pt\vrule}\hrule}}
\def\enca#1{\vbox{\hrule\hbox{
\vrule\kern8pt\vbox{\kern8pt \hbox{$\displaystyle #1$}
\kern8pt} \kern8pt\vrule}\hrule}}

\newcommand\figureframex[3]{
\begin{figure}[bth]
\hrule\hbox{\vrule\kern8pt
\vbox{\kern8pt \vbox{
\begin{center}
{\mbox{\epsfxsize=#1.truecm\epsfbox{#2}}}
\end{center}
\caption{#3}
}\kern8pt}
\kern8pt\vrule}\hrule
\end{figure}
}
\newcommand\figureframey[3]{
\begin{figure}[bth]
\hrule\hbox{\vrule\kern8pt
\vbox{\kern8pt \vbox{
\begin{center}
{\mbox{\epsfysize=#1.truecm\epsfbox{#2}}}
\end{center}
\caption{#3}
}\kern8pt}
\kern8pt\vrule}\hrule
\end{figure}
}

\newcommand{\beq}{\begin{equation}}
\newcommand{\eeq}{\end{equation}}
\newcommand{\bea}{\begin{eqnarray}}
\newcommand{\eea}{\end{eqnarray}}

\renewcommand{\and}{{\qquad {\rm and} \qquad}}

\newcommand{\virg}{{\qquad , \qquad}}

\newcommand{\tr}{{\,\rm tr}\:}

\newcommand{\Pint}{{\int\kern -1.em -\kern-.25em}}

\begin{document}
\sloppy


\pagestyle{empty}
\hfill SPhT-T04/077
\addtolength{\baselineskip}{0.20\baselineskip}
\vspace{26pt}
\begin{center}
{\large \bf {A short note about Morozov's formula}}
\end{center}
\vspace{26pt}

\begin{center}
{\sl B.\ Eynard}\hspace*{0.05cm}\footnote{ E-mail: eynard@spht.saclay.cea.fr }\\
\vspace{6pt}
Service de Physique Th\'{e}orique de Saclay,\\
F-91191 Gif-sur-Yvette Cedex, France.\\
\end{center}

\vspace{20pt}
\begin{center}
{\bf Abstract}
\end{center}

The purpose of this short note, is to rewrite Morozov's formula for correlation functions over the unitary group,
in a much simpler form, involving the computation of a single determinant.

%





\vspace{26pt}
\pagestyle{plain}
\setcounter{page}{1}


\section{Introduction}

The main result of this paper is given in \ref{mainresult}.

In \cite{morozov}, A. Morozov proposed a formula for correlation functions of unitary matrices with Itzykson-Zuber's type measure:
\beq
\langle|U_{ji}|^2\rangle_{U(N)} :=
{1\over I(X,Y)}\,\int_{U(N)}{\rm d}U  |U_{ji}|^2 {\rm
e}^{\tr(XU^\dagger YU)}
\eeq
where ${\rm d}U$ is the Haar measure over the unitary group $U(N)$ (appropriately normalized, so that \ref{IZ} below holds with prefactor $1$), and $X$ and $Y$ are two given diagonal complex matrices:
\beq
X={\rm diag}(x_1,\dots,x_N)
\virg
Y={\rm diag}(y_1,\dots,y_N)
\eeq
and the normalization factor $I(X,Y)$ is the so-called Harish-Chandra-Itzykson-Zuber integral:
\beq\label{defIZ}
I(X,Y) :=
\int_{U(N)}{\rm d}U   {\rm e}^{\tr(XU^\dagger YU)}
\eeq
It is well-known \cite{IZ, Mehta} that:
\beq\label{IZ}
I(X,Y) =
{\det E\over \Delta(X)\Delta(Y)}
\eeq
where $E$ is the matrix with entries:
\beq\label{defE}
E_{ij}:=\,{\rm e}^{x_i y_{j}}\,
\eeq
and where $\Delta(X)$ and $\Delta(Y)$ are the Vandermonde determinants:
\beq
\Delta(X):=\prod_{i<j}(x_i-x_j)
\virg
\Delta(Y):=\prod_{i<j}(y_i-y_j)\,\, .
\eeq
Here we shall assume that $\det E\neq 0$ and $\Delta(X)\neq 0$ and $\Delta(Y)\neq 0$.

\medskip

Morozov's formula was proven in \cite{mixedcorrelators}, from Shatashvili's formula \cite{shata}.
Morozov's formula was originaly written as follows, for any arbitrary sequences of complex numbers $a_i,b_j$ one has:
\bea\label{morozov}
&&\sum_{i,j=1}^N {a_i}{b_j}
\int_{U(N)}{\rm d}U  |U_{ji}|^2 {\rm
e}^{\tr(XU^\dagger YU)} =  \frac 1{\Delta(X)\Delta(Y)}
\sum_{\rho\in S_N} \epsilon(\rho) {\rm e}^{\sum x_\ell y_{\rho(\ell)}} \cr
&& \sum_{n=0}^{N-1} (-1)^{n}
\sum_{i_1<i_2<\dots<i_{n+1}}
\hspace{-20pt}
 \frac {
\det\pmatrix{
1&\cdots&1\cr
x_{i_1}& \cdots& x_{i_{n+1}}\cr
\vdots& &\vdots\cr
x_{i_1}^{n-1}& \cdots & x_{i_{n+1}}^{n-1}\cr
a_{i_1} & \cdots & a_{i_{n+1}}}}{
\det\pmatrix{
1&\cdots&1\cr
x_{i_1}& \cdots& x_{i_{n+1}}\cr
\vdots& &\vdots\cr
x_{i_1}^{n}& \cdots & x_{i_{n+1}}^{n}}}
\frac {
\det\pmatrix{
1&\cdots&1\cr
y_{\rho(i_1)}& \cdots& y_{\rho(i_{n+1})}\cr
\vdots& &\vdots\cr
y_{\rho(i_1)}^{n-1}& \cdots & y_{\rho(i_{n+1})}^{n-1}\cr
b_{\rho(i_1)} & \cdots & b_{\rho(i_{n+1})}}}{
\det\pmatrix{
1&\cdots&1\cr
y_{\rho(i_1)}& \cdots& y_{\rho(i_{n+1})}\cr
\vdots& &\vdots\cr
y_{\rho(i_1)}^{n}& \cdots & y_{\rho(i_{n+1}) }^{n}}}\ ,\label{moroformula} \cr
\eea
where $S_N$ is the smmetric group of rank $N$.

\section{Rewriting Morozov's formula}

The purpose of this short note is to rewrite this formula in a much simpler form, in a way very similar to what was done in \cite{mixedcorrelators}.

\medskip

For any two complex numbers $x$ and $y$ (such that $|x|>{\rm max}\, |x_i|$ and $|y|>{\rm max}\, |y_i|$), consider the choice:
\beq
a_i={1\over x-x_i}=\sum_{r=0}^\infty {x_i^r\over x^{r+1}}
\virg
b_i={1\over y-y_i}=\sum_{s=0}^\infty {y_i^s\over y^{s+1}}
\eeq

Introduce the Schur polynomials (corresponding to hook diagramms):
\bea
S_r(x_{i_1},\dots,x_{i_{n+1}})  &:=&
\frac {
\det\pmatrix{
1&\cdots&1\cr
x_{i_1}& \cdots& x_{i_{n+1}}\cr
\vdots& &\vdots\cr
x_{i_1}^{n-1}& \cdots & x_{i_{n+1}}^{n-1}\cr
x_{i_1}^r & \cdots & x_{i_{n+1}}^{r}}}{
\det\pmatrix{
1&\cdots&1\cr
x_{i_1}& \cdots& x_{i_{n+1}}\cr
\vdots& &\vdots\cr
x_{i_1}^{n}& \cdots & x_{i_{n+1}}^{n}}} \cr
&=&  \sum_{a_1\leq
a_2\leq\dots\leq a_{r-n}} \prod_{k=1}^{r-n} x_{i_{a_k}}
=
 \sum_{j_1+\cdots+j_{n+1}=r-n} x_{i_1}^{j_1}\cdots
 x_{i_{n+1}}^{j_{n+1}}\label{hook}\  .
\eea

The formal generating function of these Schur polynomials is:
\beq\label{Schurgenerating}
\sum_{r=0}^\infty {1\over x^{r+1}}S_r(x_{i_1},\dots,x_{i_{n+1}}) = \prod_{k=1}^{n+1} {1\over x-x_{i_k}}
\eeq

Inserting that into \ref{morozov}, one gets:
\bea
&&\hspace{-1.6cm}I(X,Y)\,\sum_{i,j}
{1\over x-x_i}\,{1\over y-y_j}\, \langle|U_{ji}|^2\rangle_{U(N)} \cr
&& \hspace{-1.5cm}	=\int_{U(N)}{\rm d}U  \,\,
\tr \left({1\over x-X}U{1\over y-Y}U^\dagger\right)\,
{\rm e}^{\tr(XU^\dagger YU)}  \cr
&& \hspace{-1.5cm}=\frac 1{\Delta(X)\Delta(Y)}
\sum_{r,s=0}^\infty \sum_{\rho\in S_N} \epsilon(\rho) {\rm e}^{\sum x_\ell y_{\rho(\ell)}} \cr
&& \sum_{n=0}^{N-1} (-1)^{n}\!\!\!\!\!\!\!\!\!\!
\sum_{i_1<i_2<\dots<i_{n+1}}
{S_r(x_{i_1},\dots,x_{i_{n+1}})\over x^{r+1}}
\,\,
{S_s(y_{\rho(i_1)},\dots,y_{\rho(i_{n+1})})\over y^{s+1}}
 \cr
&& \hspace{-1.5cm}=\frac 1{\Delta(X)\Delta(Y)}
 \sum_{\rho\in S_N} \epsilon(\rho) {\rm e}^{\sum x_\ell y_{\rho(\ell)}}
\sum_{n=0}^{N-1} (-1)^{n}\!\!\!\!\!\!\!\!\!\!
\sum_{i_1<i_2<\dots<i_{n+1}}
\prod_{k=1}^{n+1} {1\over x-x_{i_k}}
\,\,
\prod_{l=1}^{n+1} {1\over y-y_{\rho(i_l)}}
 \cr
&& \hspace{-1.5cm}=\frac 1{\Delta(X)\Delta(Y)}
 \sum_{\rho\in S_N} \epsilon(\rho) {\rm e}^{\sum x_\ell y_{\rho(\ell)}}
\left[
1-\prod_{i=1}^{N} \left(1-{1\over x-x_{i}}{1\over y-y_{\rho(i)}}\right)
\right]
 \cr
&& \hspace{-1.5cm}=\frac 1{\Delta(X)\Delta(Y)}
 \sum_{\rho\in S_N} \epsilon(\rho)
\left[\prod_{i=1}^{N}\,{\rm e}^{x_i y_{\rho(i)}}\,
-\prod_{i=1}^{N} \left(\,{\rm e}^{x_i y_{\rho(i)}}\,
-{1\over x-x_{i}}
\,{\rm e}^{x_i y_{\rho(i)}}\,
{1\over y-y_{\rho(i)}}\right)\right]
 \cr
&& \hspace{-1.5cm}=\frac 1{\Delta(X)\Delta(Y)}
\left[
\det\left(\,{\rm e}^{x_i y_{j}}\,\right)
-\det\left(\,{\rm e}^{x_i y_{j}}\,-
{1\over x-x_{i}}
\,{\rm e}^{x_i y_{j}}\,
{1\over y-y_{j}}\right)
\right]
 \cr
\eea

Thus, Morozov's formula can be rewritten:
\beq
\int_{U(N)}{\rm d}U  \,\,
\tr \left({1\over x-X}U{1\over y-Y}U^\dagger\right)\,
{\rm e}^{\tr(XU^\dagger YU)}
=
{\det E
-\det \left(E-{1\over x-X}E{1\over y-Y}\right)
\over \Delta(X)\Delta(Y)}
\eeq
or:
\beq\label{mainresult}
\encadremath{
\begin{array}{rcl}
{\displaystyle  \left<\tr \left({1\over x-X}U{1\over y-Y}U^\dagger\right)\, \right>}
& = &
{\displaystyle 1-{\det \left(E-{1\over x-X}E{1\over y-Y}\right)\over \det E} }\cr
& = &
{\displaystyle 1- \det \left(1-{1\over x-X}E{1\over y-Y}E^{-1}\right)}
\end{array}}
\eeq

\section{Concluding remarks}

From that expression of Morozov's formula, it is rather easy to recover any individual correlator by taking residues:
\beq
 \left< U_{ij}\, U^\dagger_{ji} \right>
 =
 \mathop{\rm Res}_{x\to x_i}
 \mathop{\rm Res}_{y\to x_j}
 \left<\tr \left({1\over x-X}U{1\over y-Y}U^\dagger\right)\, \right>
\eeq

\medskip

Notice also that \ref{mainresult} is very similar to what was found in \cite{mixedcorrelators} after integration over $X$ and $Y$.

\medskip

Notice that by expanding the determinant in \ref{mainresult} along its last column, one can find a recursion relation relating
$U(N)$ to $U(N-1)$ integrals, which is equivalent to Shatashvili's approach \cite{shata}.

\bigskip

{\em Aknowledgements:} the author wants to thank M. Bertola and J. Harnad for discussions about that topic.
The author thinks that this short note should be seen as an addendum to the article \cite{mixedcorrelators}.


%
%

\end{document}